# Zinc Oxide Modified with Benzylphosphonic Acids as Transparent Electrodes in Regular and Inverted Organic Solar Cell Structures

Ilja Lange[1], Sina Reiter[1], Juliane Kniepert[1], Fortunato Piersimoni[1], Michael Pätzel[2], Jana Hildebrandt[2], Thomas Brenner[1], Stefan Hecht[2], Dieter Neher[1,a)]

[1] *Institute of Physics and Astronomy, University of Potsdam, Karl-Liebknecht-Strasse 24-25, 14476 Potsdam, Germany*

[2] *Department of Chemistry & IRIS Adlershof, Humboldt-Universität zu Berlin, Brook-Taylor-Str. 2, 12489 Berlin, Germany*

*ABSTRACT*

An approach is presented to modify the WF of solution-processed sol-gel derived ZnO over an exceptionally wide range of more than 2.3 eV. This approach relies on the formation of dense and homogeneous self-assembled monolayers based on phosphonic acids with different dipole moments. This allows us to apply ZnO as charge selective bottom electrodes in either regular or inverted solar cell structures, using P3HT:PCBM as the active layer. These devices compete with or even exceed the performance of the reference cell on ITO/PEDOT:PSS. Our finding challenges the current view that bottom electrodes in inverted solar cells need to be electron-blocking for good device performance.

Transparent metal oxides (TMOs) are integral parts of today's optoelectronic devices, either as electron conducting electrodes or intrinsically-doped semiconductors. Among these, zinc oxide (ZnO) is becoming increasingly important, as it is composed of earth-abundant elements. Also, a wide range of vapor- and solution based methods is available for ZnO deposition, ranging from chemical or physical vapor deposition (CVD, PVD) for the preparation of epitaxially grown single crystalline ZnO films to sol-gel procedures for low cost solution-based deposition. ZnO is already intensively used as a low-cost, transparent electrode in inorganic devices,[1-2] but more recently also as a charge selective interlayer material in efficient organic solar cells (OSCs) or organic light-emitting diodes (OLEDs).[3-4] However, the moderate work function (WF) of untreated ZnO of about 4.3 eV causes injection barriers to almost all conventional organic semiconductors[5-9] and the WF of such ZnO layers is poorly reproducible due to the physisorption of contaminations.[10] Finally, its poor chemical stability in an acidic environment [11-13] makes the application particularly in organic devices challenging. Therefore, for some of the recent organic record solar cells, the WF of the ZnO forming the bottom electrode was

---

a) Author to whom correspondence should be addressed. Electronic mail: neher@uni-potsdam.de

lowered by applying an electrolytic (ionic) buffer layer such as PEI or PEIE.[14] However, the chemical and physical structure of the ZnO/PEI/OSC interface is still not understood.

An elegant way to alter the WF and, at the same time, passivate the ZnO surface is the attachment of a self-assembled monolayer (SAM) of polar molecules. SAM modifications have already been applied successfully to other transparent metal oxides such as ITO,[15-17] however, deposition of these mostly acidic molecules onto ZnO is well-known to cause etching of the chemically more unstable ZnO surface.[18-19] Therefore, recent attempts drastically reduced the exposition time to the acidic environment simply by spin-coating the solution of the SAM forming molecules onto ZnO.[20-21] However, given the short time for molecular assembly time, the quality of the so-formed monolayer might be ill defined. Indeed, SAM formation via spincoating yielded a rather small shift of the WF, indicating a sparse coverage.

Recently we succeeded to relate the degree of etching of the ZnO surface upon SAM preparation to the water content in the molecular solution (and therefore a reduced protonation of the PA molecules).[5] Experiments on well-defined single crystalline ZnO (0001) and (000-1) surfaces proved that etching during immersion can be fully suppressed if explicitly dried ethanol (<0.01% $H_2O$) is used as solvent. A preparation protocol was developed which results in a very dense and homogenous monolayer of oriented molecules on ZnO substrates. The resulting SAMs were characterized by a range of complementary techniques, revealing the coverage of the ZnO surface by a dense monolayer with a tilting angle α of the aromatic ring with respect to the ZnO surface of 47° ± 3°, which is in perfect agreement to previous theoretical predictions on that system in case of tridental binding of the phosphonic acid anchoring group to the substrate.[22] By employing substituted benzylphosphonic acids (BPA) as well as pyrimidin phosphonic acid (PyPA) with different dipole moments of the head group, we succeeded to alter the WF (measured with Kelvin Probe (KP) inside a glovebox) from about 4.1 eV to almost 5.7 eV. A strict linear relation between the dipole moment and the resulting WF was thereby observed, which is in agreement to the prediction of Helmholtz equation. From the analysis of this linear relation, a molecular density of ~2 $nm^{-2}$ was concluded which is close to the theoretical maximum of 2.73 $nm^{-2}$.[22] Atomic force microscopy (AFM) and scanning Kelvin probe microscopy (SKPM) revealed a very homogenous SAM structure at least at the few 10 nm scale. In addition, the wide achievable WF range allowed us already to apply the modified ZnO as highly selective electron-injecting as well as hole-injecting contact in electron- or hole-only diodes, simply by varying the used SAM molecule, proving the electronic grade of the so prepared SAM/ZnO electrodes.

Here, we extend these investigations to the commercially more relevant, though structurally less defined, polycrystalline surface of sol-gel processed ZnO. We are able to tune the WF of the this sol-gel-processed ZnO over an exceptional wide range of more than 2.3 eV, enabling the application as efficient transparent cathode in inverted P3HT:PCBM solar cell structures and, for the first time, as efficient anode also in regular structures. Thereby, all devices with SAM-modified ZnO electrodes compete with or even exceed the performance of the regular benchmark system comprising an ITO/PEDOT:PSS electrode, highlighting again the potential of the method.

As ZnO substrates either (10-10) single crystalls (CrysTec GmbH, Germany) or sol-gel-processed ZnO was used. The sol-gel-ZnO films were prepared according to literature[4] via spincoating the solution comprising 100 mg zinc acetate dehydrate dissolved in 1 ml 2-methoxyethanol with 27.7 µl ethanolamine onto ITO-covered glass, yielding a ~30 nm thick layer which was subsequently annealed at 200°C for one hour to promote crystallization. PA molecules for SAM formation (for chemical names and supplier see Ref. [5]) were dissolved in explicitly dried ethanol (SeccoSolv®, Merck Millipore, <0.01% $H_2O$) in a concentration of 2 mM inside a glovebox to prevent water adsorption. Preparation of the SAMs was performed by immersing the ZnO substrates into this solution on a hot plate at 70°C in covered glass vessels, subsequent gentle purging in ethanol and drying at 90°C for 3 min on a hot plate. For solar cells, P3HT:PCBM (Riecke Metals Inc./ Solenne) bulk heterojunction layers in a 1:1 (wt./wt.) ratio were prepared by spincoating from a chloroform solution and annealed at 150°C for 15 min at a hot plate. 10 nm Sm/ 100 nm Al (regular) or 7 nm $MoO_3$/ 100 nm Ag (inverted) top electrodes were evaporated on top leading to pixel sizes of 6 $mm^2$.

The three most stable facets of a ZnO crystal are the two polar (0001)-Zn and (000-1)-O as well as the mixed terminated (10-10) surface.[23] It can be expected that also the surface of such sol-gel-processed ZnO layers consists mainly of these three facets. We, therefore, first tested our approach to the mixed terminated (10-10) single crystalline surface. Again, annealing the samples at 1180 °C for 6 h precipitated a well-defined smooth terraced surface (see Figure 1a). Following the protocol of Ref. [5], SAMs from various BPAs and PyPA (see Figure 2a) were prepared on top of these ZnO substrates. We again find that the WF shift is largest for a specific optimum of immersion time, between 20 min and 2 h. Smaller WF shifts for non-optimized immersion are supposedly due to an incomplete coverage for shorter times or stable multilayer formation at longer times. The absolute WF as function of the molecular head-group dipole moment of the mixed terminated ZnO surface (Figure 2b) agrees thereby nicely with results obtained on the two modified polar surfaces, resulting in equal slopes in that diagram. Kedem et al. recently also found in their experiments that the particular choice of the ZnO substrate has only little effect on the resulting WF upon phenylphosphonate-treatment.[24] Although the (10-10) ZnO surface differs geometrically from the two polar surfaces and supports a rather bidental binding of the PA, we conclude that the effects on surface coverage and molecular tilt are negligible. Furthermore, our AFM and SKPM measurements revealed also on the (10-10) ZnO surface a very homogenous coverage with a small variation of the local WF and without etching effects (Figure 1a-c). We can therefore conclude that a similarly high SAM quality is obtained also on the mixed terminated (10-10) ZnO surface, notably with the same preparation parameters as used for the two polar surfaces.

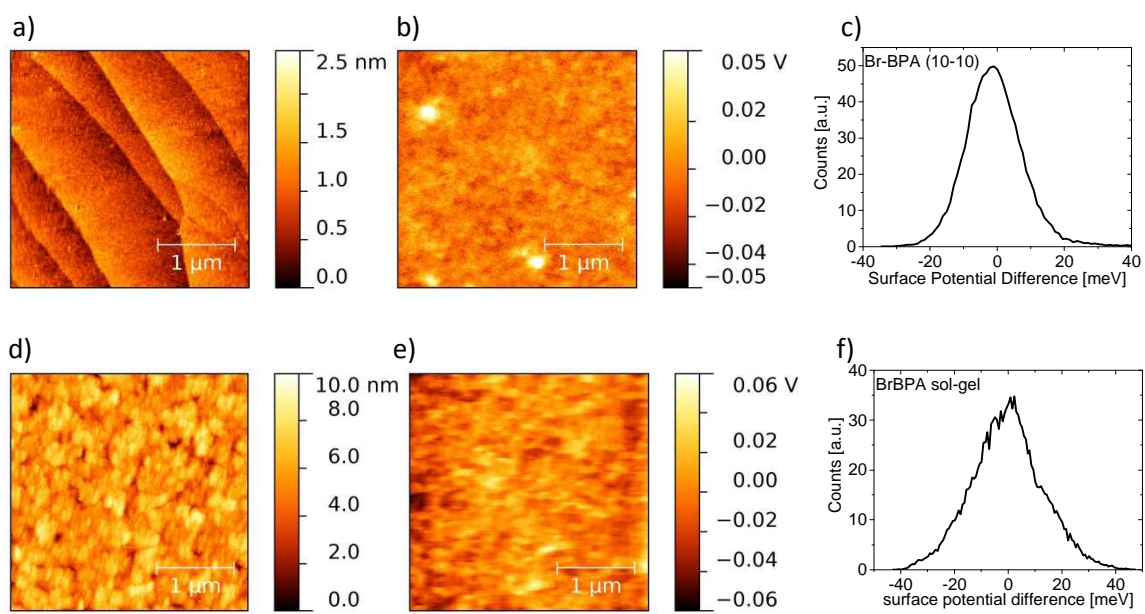

**Figure 1.** AFM (a,d), SKPM (b,e) and SKPM histogram (c,f) of BrBPA-SAM covered single crystalline (10-10) (a-c) or polycrystalline sol-gel-processed (d-f) ZnO substrates.

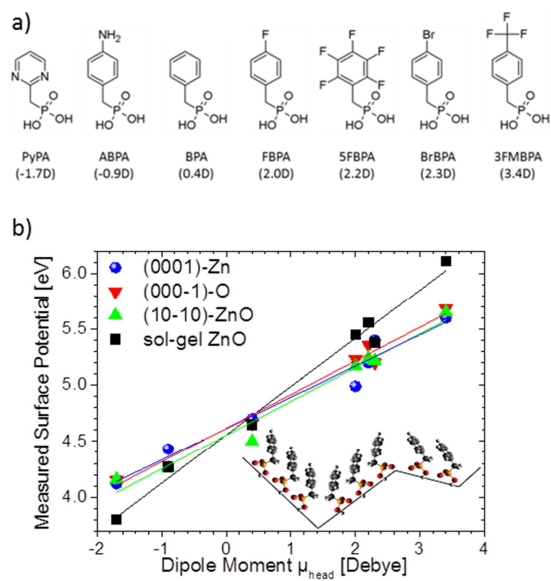

**Figure 2.** a) Chemical structures of the applied SAM-forming phosphonic acids with their respective molecular dipole moment of the head group in brackets. b) Measured surface potential (KP) as function of the molecular head group dipole moment of the three single crystalline and the sol-gel-processed ZnO surfaces covered with the molecular SAMs displayed above. The strict linear relation agrees to Helmholtz equation. The slightly higher slope of the sol-gel-processed ZnO is most likely caused by a preferential upright orientation of the molecular head group on the rougher surface structure (inset).

Having established a preparation protocol for PA SAMs on the three important ZnO crystal surfaces, we then turned to modifying the polycrystalline surface of sol-gel prepared ZnO. Notably, to reach a maximum WF shift of the sol-gel ZnO immersion times between 20 min and 2 h were needed which is equivalent to what was used for the single crystalline surfaces, albeit with no significant further change for even longer immersion times. However, for ideal electronic behavior it turned out that the immersion time needed to be increased to about 4 h, presumably due to a filling/ passivation of holes in the porous ZnO layer. The resulting WFs as function of the dipole moment of the SAM molecules head group are also shown in Figure 2. Notably, comparable or even stronger WF shifts were observed on these polycrystalline sol-gel-prepared substrates, covering an exceptionally wide WF range from 3.8 eV to above 6.1 eV.

According to the Helmholtz equation the WF shift is proportional to the dipole moment component perpendicular to the surface $\mu_\perp = \mu \cdot \sin\alpha$ and to the number of dipoles per area $N/A$

$$\Delta WF = \frac{e \cdot \mu \cdot \sin\alpha}{\varepsilon_0 \cdot k_{red}} \cdot \frac{N}{A}, \tag{1}$$

where $e$ denotes the elementary charge and $\varepsilon_0$ is the vacuum permittivity. $k_{red}$ accounts for the depolarization of the molecules in close proximity and is about 2 in this system.[25] A plausible explanation for the higher WF shifts arises from the fact that the sol-gel-prepared ZnO has a significantly larger surface roughness compared to the terraced, well-defined single crystalline surface (see Figure 1). Such a larger surface roughness causes a larger effective surface area, supporting more binding sites for the PAs; however, a tilt of the local surface reduces also $\mu_\perp$ in the same quantity. We propose instead a non-isotropic distribution of the molecular tilt resulting in a more upright orientation of the head group with respect to the surface plane on this particular rough surface (see inset in Fig. 2b). Notably, the SKPM measurements in Figure 1 reveal an only minor increase of the lateral WF variations after SAM-modification compared to the single-crystalline ZnO surface. We therefore conclude that a well-defined, high quality SAM was also formed on the sol-gel-ZnO.

With the WF being tunable over a wide range, these modified sol-gel processed ZnO films can be applied as an efficient transparent cathode in inverted solar cells, but also as the transparent anode in regular P3HT:PCBM solar cell structures without the need of additional buffer layers such as PEDOT:PSS. For the inverted structure, the sol-gel ZnO surface was modified by an amino-BPA (ABPA) SAM, yielding a WF of about 4.25 eV. In addition, 3-fluoromethyl (3FM)-BPA was used to prepare a ZnO-based anode with a WF of about 6.1 eV. A benchmark reference cell on ITO/PEDOT:PSS in a regular structure was also prepared for comparison.

The resulting *J-V*-curves under illumination are shown in Figure 3 with the respective performance parameters in Table I. All solar cells comprising ZnO as bottom electrode show an exceptionally good performance, with a power conversion efficiency (PCE) of the inverted structure distinctly exceeding the reference value of the benchmark cell on ITO/PEDOT:PSS. Notably, also the regular solar cell comprising the 3FMBPA-SAM modified ZnO as the anode displays a high PCE, comparable to that of the ITO/PEDOT:PSS reference cell, highlighting again the electronic grade of our SAM-modified ZnO electrodes. As a general trend, all ZnO-based devices benefit from a higher short circuit current ($J_{SC}$), however, this effect is most distinct in the inverted structure with an almost 20% higher $J_{SC}$ with respect to the reference cell. The origin of such an increase was recently attributed to a more ideal optical field profile in inverted solar cells.[26] Other possible effects will be addressed below. On the other hand, both the fill factor (*FF*) and the open circuit voltage ($V_{OC}$) were slightly lower than for the reference cell. It has been proposed that the good performance of cells with PEDOT:PSS is due to the electron-blocking nature of the conducting polymer composite.[27] Here, replacing PEDOT:PSS with the 3FM-BPA-modified ZnO has an only weak effect on the overall photovoltaic characteristics, though the modified ZnO is not expected to serve as an electron-blocking layer. Our result therefore challenges the current view that bottom electrodes in inverted solar cells need to be electron-blocking for good performance. In agreement to this finding, recent device simulations revealed a negligible effect of surface recombination on the *J-V*-characteristics of P3HT:PCBM devices.[28]

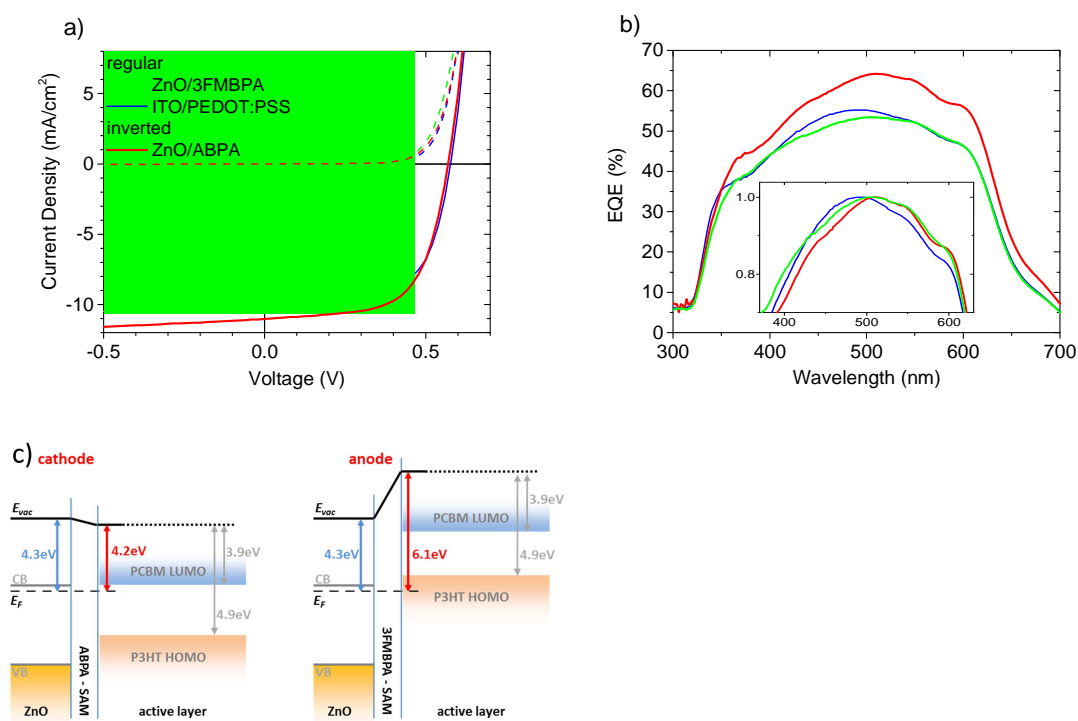

**Figure 3.** a) *J-V*-characteristics of all P3HT:PCBM solar cells under illumination (full line) and in the dark (dashed line). b) EQE spectra of the respective devices with the normalized spectra displayed in the inset. c) Energy scheme of the contact between the SAM-modified ZnO used as cathode (left) or anode (right) and the P3HT:PCBM blend.

**Table I.** Device parameters of all solar cells.

| Bottom electrode | ECE (%) | $V_{OC}$ (V) | $J_{SC}$ (mA/cm$^2$) | FF (%) |
|---|---|---|---|---|
| regular | | | | |
| ZnO/3FMBPA SAM | 3.82 | 0.560 | 9.82 | 63.4 |
| ITO/PEDOT:PSS | 3.97 | 0.578 | 9.34 | 67.9 |
| inverted | | | | |
| ZnO/ABPA SAM | 4.22 | 0.570 | 11.0 | 63.3 |

The overall increase of $J_{SC}$ in regular or inverted solar cells prepared on ZnO electrodes goes along with corresponding changes in the EQE spectra as shown in Figure 3b. One reason for this improvement might be the higher transparency of ZnO, allowing more light to reach the active layer. However, the normalized EQE spectra shown in the inset of Figure 3b also reveals a more pronounced shoulder at 550 nm and 600 nm, which points to a more crystalline P3HT phase in the blends coated on ZnO. This finding is, in turn, consistent with the lower $V_{OC}$ measured on some of the ZnO-based devices, as a higher degree of crystallinity of the polymer in P3HT:PCBM blends is known to reduce the effective band gap between the donor-HOMO and acceptor-LUMO.[29]

In conclusion we have successfully modified the WF of solution-processed sol-gel derived ZnO by employing BPA-based SAMs with different dipole moments. For this purpose, a recently developed protocol for the formation of SAMs on highly defined single-crystalline ZnO was adopted to the less defined surfaces of sol-gel processed ZnO. This modification caused the WF to vary over an exceptionally large range of more than 2.3 eV, which enabled us to apply ZnO, for the first time, also as hole injecting/extraction interlayer in regular solar cell structures. Efficiency measurements of regular and inverted P3HT:PCBM solar cells showed that these fully solution processed interlayers lead to a competitive or even increased device performance, compared to the reference cells on ITO/PEDOT:PSS. As these SAM modified ZnO layers can be coated via solution-based methods on virtually any electrode material, the presented approach paves the way to ITO-free devices without the need for including electrolytic buffer layers. Moreover, the tunability from cathode to anode opens novel degrees of freedom in device design but also for a wider understanding of the role of the WF on injection and extraction of charges.

ACKNOWLEDGMENT

This work was financially supported by the Deutsche Forschungsgemeinschaft (DFG) within the Collaborative Research Centre HIOS (SFB 951) and the SPP 1355,